\documentclass[12pt,preprint]{aastex}

\newcommand{\msun}{$M_{\odot}$}
\newcommand{\vmon}{A0620-00}
\newcommand{\sj}{Swift J1753.5-0127}
\newcommand{\flam}{ergs~cm$^{-2}$~s$^{-1}$~\AA$^{-1}$}

\begin{document}
\title{Multiwavelength Observations of \sj}

\author{Cynthia S. Froning\altaffilmark{1,2}}
\email{cynthia.froning@colorado.edu}
\affil{Center for Astrophysics and Space Astronomy, University of Colorado, \\
  593 UCB, Boulder, CO 80309-0593}
  
\author{Thomas J. Maccarone}
\email{t.j.maccarone@soton.ac.uk}
\affil{Department of Physics, Texas Tech University, Box 41051, Lubbock, TX 79409}
        
\author{Kevin France}
\email{kevin.france@colorado.edu}
\affil{Center for Astrophysics and Space Astronomy, University of Colorado, \\
  593 UCB, Boulder, CO 80309-0593}
      
\author{Lisa Winter}
\email{lwinter@aer.com}
\affil{Space Weather and Effects Division, Atmospheric and Environmental Research, Superior, CO}        
\author{Edward L. Robinson}
\email{elr@astro.as.utexas.edu}
\affil{Department of Astronomy, University of Texas at Austin, 
Austin, TX 78712}

\author{Robert I. Hynes}
\email{rih@phys.lsu.edu}
\affil{Department of Physics and Astronomy, Louisiana State University, 
Baton Rouge, LA 70803}

\and
 
\author{Fraser Lewis}
\email{flewis@glam.ac.uk}
\affil{Faulkes Telescope Project, University of South Wales; Astrophysics Research Institute, Liverpool John Moores University}
\altaffiltext{1}{Department of Astrophysical and Planetary Sciences, University of Colorado}
\altaffiltext{2}{University Affiliate Research Fellow, Department of Astronomy, University of Texas at Austin}

\begin{abstract}
We present contemporaneous X-ray, ultraviolet, optical and near-infrared observations of the black hole binary system, \sj, acquired in 2012 October.  The UV observations, obtained with the Cosmic Origins Spectrograph on the Hubble Space Telescope, are the first UV spectra of this system. The dereddened UV spectrum is characterized by a smooth, blue continuum and broad emission lines of \ion{C}{4} and \ion{He}{2}. The system was stable in the UV to $<$10\% during our observations. We estimated the interstellar reddening by fitting the 2175~\AA\ absorption feature and fit the interstellar absorption profile of Ly$\alpha$ to directly measure the neutral hydrogen column density along the line of sight. By comparing the UV continuum flux to steady-state thin accretion disk models, we determined upper limits on the distance to the system as a function of black hole mass. The continuum is well fit with disk models dominated by viscous  heating rather than irradiation. The broadband spectral energy distribution shows the system has declined at all wavelengths since previous broadband observations in 2005 and 2007. If we assume that the UV emission is dominated by the accretion disk the inner radius of the disk must be truncated at radii above the ISCO to be consistent with the X-ray flux, requiring significant mass loss from outflows and/or energy loss via advection into the black hole to maintain energy balance.
\end{abstract}


\section{Introduction} \label{sec_intro}

X-ray binaries (XRBs) are interacting binary systems in which a black hole or neutron star accretes material from a donor star, typically via an accretion disk. XRBs can occupy several different accretion states defined by their X-ray activity \citep{mcclintock2006}.  Systems in the low state have luminosities at or below several percent of their Eddington luminosities, although due to hysteresis effects in the transitions between states, there is not a strict monotonic relationship between luminosity and spectral state \citep{maccarone2003,maccarone2003b}. The low (or ``low/hard") state systems are also characterized by hard, non-thermal X-ray spectra fit with power-law components with photon indices $\sim$1.7. This is generally ascribed to thermal Comptonization in a hot, optically thin corona located near the black hole or neutron star \citep{thorne1975,shapiro1976,sunyaev1979}, although the source and physical properties of the corona are controversial. In contrast, XRBs in the high (or ``high/soft") state have higher luminosities and a strong thermal component to the soft X-ray spectrum. The thermal component is interpreted as originating in the accretion disk and is fit by models of multi-temperature emission from a steady-state, thin disk, where the disk temperature varies radially as T(R) $\propto$ R$^{-3/4}$ \citep{shakura1973}. XRBs in the low state are also host to persistent collimated outflows. Flat or inverted spectra have been observed in the radio to infrared regime in most XRBs in the low state, similar to the signature emission of jet cores in AGN \citep{jain2001,fender2006,russell2006}. The spectra are attributed to self-absorbed synchrotron emission from a compact, highly collimated outflow that is quenched when a system transitions from the low to the high state \citep{tananbaum1972,fender1999}.  

While great progress has been made in observing and modeling the behavior of XRBs, a number of questions remain unresolved, including the physical structure of the accretion disk, the link between accretion and outflows (in the forms of jets and winds), and the relative contributions of these components to the broadband spectral energy distributions (SEDs). \citet{esin1997} presented a unified model of the accretion states of XRBs within the paradigm of advection dominated accretion flows (ADAFs). In the ADAF model picture, a thin accretion disk in the low state is disrupted outside the innermost stable circular orbit (ISCO) of the black hole, forming a spherical, radiatively inefficient corona \citep{narayan1995}. Because the corona is radiatively inefficient, most of the accretion energy released by viscous dissipation is advected with the flow into the black hole. As a system transitions to the high state in outburst, the ADAF disappears and the thin disk extends to the ISCO. However, other models of the source of the X-ray corona in the low state have been advanced, including coronae driven by magnetic flares from the underlying accretion disk or emission from the base of the jet itself \citep{haart1991,merloni2001a,merloni2001b,markoff2001}. 

Multiwavelength observations provide powerful tools for studying the structure of the disks and outflows in XRBs. The ultraviolet (UV) is a key component of these observations. UV continuum data probe the expected peak temperatures of thermal emission components in quiescent disks, while UV line emission traces the kinematics and abundances of accretion disk chromospheres \citep[e.g.,][]{froning2011,bayless2010, hynes2009,haswell2002,mcclintock2000}. The UV can also directly test competing models of the structure of XRBs in the low state. In 2010, we obtained far-ultraviolet (FUV) observations of the black hole XRB, \vmon\ \citep{froning2011}. The FUV component of the spectral energy distribution proved to be key to constraining accretion disk and jet models. Contrary to previous observations, the optical-UV spectrum did not continue to drop to shorter wavelengths, but instead showed a recovery and an increasingly blue spectrum in the FUV. The blue FUV trend, combined with the strong UV variability between observing epochs, was inconsistent with the previous ADAF model for \vmon\ \citep{narayan1997} and may indicate significant changes in the size of the ADAF and the thin disk/ADAF transition radius over time. Alternately, a comparison of the SED with a jet-dominated model  showed good agreement and predicted that the increasing FUV flux originates in emission from the pre-acceleration inner jet component of the outflow. 

In this manuscript, we continue our investigation of the FUV properties of XRBs and present the first FUV observations of \sj, obtained using the Cosmic Origins Spectrograph (COS) aboard the Hubble Space Telescope \citep{green2012}. \sj\ is a X-ray transient discovered by the Swift Burst Alert Telescope when it went into outburst in 2005 \citep{palmer2005}. It was soon detected in the UV, optical, near-infrared, and radio bands as well \citep{still2005,halpern2005,torres2005,fender2005}. The shape of the hard X-ray spectrum was consistent with those of black hole X-ray binaries in the hard state \citep{cadolle2005}. 
During its outburst, \sj\ remained in the low state rather than transitioning to the high state, a behavior seen in other low-luminosity XRBs, including XTE J1118+480 and Swift J1357.2-0933 \citep{brocksopp2004,hynes2000,armas2013}. 
%Like these sources, \sj\ is located above the disk plane and is a likely Galactic halo source \citep{cadolle2007}. 

The system has not returned to quiescence since its 2005 outburst, remaining active at X-ray and optical/infrared wavelengths \citep[e.g.,][]{shaw2013}. Numerous fits to the X-ray spectrum of \sj\ have differed in their conclusions concerning whether a thermal accretion disk component is present in the spectrum and whether the inner disk radius is located at the ISCO or at a larger radius \citep{miller2006, ramadevi2007,reis2009,hiemstra2009,reynolds2010,mostafa2013}. Here, we examine the broadband SED of \sj\ seven years after its initial outburst using COS observations combined with X-ray, optical, and near-infrared data obtained on the same day. In \S~\ref{sec_observ} we present the multiwavelength observations and data reduction; \S~\ref{sec_analysis} shows the analysis of the UV spectrum and the SED; and in \S~\ref{sec_discussion} and \S~\ref{sec_conclusions} we discuss the results and the conclusions from our work.

\section{Observations and Data Reduction} \label{sec_observ}

On 2012 October 02 we obtained X-ray, UV, optical, and near-infrared (NIR) observations of \sj. The observations are summarized in Table~\ref{tab_obs}. The X-ray and UV observations were simultaneous, while the optical/NIR data were acquired several hours earlier. 

\subsection{HST UV Spectroscopy}

We observed \sj\ with COS on 2012 Oct 02. Information about the design and on-orbit performance of COS can be found in \citet{osterman2011}, \citet{green2012} and in the COS Instrument Handbook \citep{holland2012}. We used the G130M and G160M gratings to obtain the FUV (1150 -- 1800~\AA) spectrum at moderate spectral resolution ($R \simeq$~18,000). We stepped each grating through four central wavelength positions to minimize the effects of fixed pattern noise. We also obtained NUV spectra using the G230L grating taken in two grating tilt positions (the 2950~\AA\ and 3360~\AA\ settings). Each G230L exposure covers three non-contiguous spectral regions, so the two tilt positions together subtend 1650 -- 3560~\AA\ with some gaps. The spectral resolution of the G230L data is $R \simeq$~ 3000.

We retrieved the COS data from the Multi-Mission Archive at STScI (MAST).  The data had been processed with V. 2.18.5 of CALCOS. For the FUV data, we coadded the output one-dimensional spectral data products using a custom IDL code, described in \citet{danforth2010}\footnote{See also \url{http://casa.colorado.edu/~danforth/science/cos/costools.html}}.  The code combines different wavelength settings and creates a weighted mean spectrum (with lower weight given to regions of uncertain flux calibration near detector edges). For the NUV spectra we used the X1D files as delivered by CALCOS.

\subsection{Swift X-ray and UV Imaging}

Swift acquired pointed observations of \sj\ for 2100 sec on 2012 October 02, simultaneous with the HST/COS observations. We retrieved and calibrated the X-Ray Telescope (XRT; Burrows et al. 2005) and UV/Optical Telescope (UVOT; Roming et al. 2005) data.  The XRT was in windowed timing mode for the observation.  A total of 12355 X-ray counts were detected. The background is $\simeq$20 photons, so a background subtraction in the spectrum was neglected during the analysis. For the UVOT data, we used the Swift FTOOL\footnote{See  \url{http://heasarc.gsfc.nasa.gov/docs/software/ftools/ftools\_menu.html} } uvotsource to extract the background-subtracted flux from the UVOT observations in the U filter. The U-band magnitude is given in Table~\ref{tab_oir}.  

For the XRT observation, we fit the data with an absorbed power law model \citep[phabs*powerlaw in XSPEC version 12.0;][]{arnaud1996}. If the foreground absorption is allowed to float freely we find $N_H = 2.66\pm0.14\times10^{21}$ cm$^{-2}$ and $\Gamma=2.00\pm0.04$, with the errors assumed to be purely statistical, and the uncertainties given at the 90\% confidence level. The fit has $\chi^2_\nu= 1.01$ (with 769 degrees of freedom). The absorbed flux in the 0.3--8 keV band is $2.37\times10^{-10}$~ergs~cm$^{-2}$~s$^{-1}$, and the unabsorbed flux in the same band is $3.8\times10^{-10}$~ergs~cm$^{-2}$~s$^{-1}$. If we fix the interstellar hydrogen absorption at the value found by our fits to the Ly$\alpha$ absorption line profile ($N_H = 2.0\times10^{21}$ cm$^{-2}$; see \S~\ref{sec_ebv}) we obtain $\Gamma=1.81\pm0.02$ and $\chi^2_\nu= 1.11$ (768 $dof$). The 0.3 -- 8.0 keV unabsorbed flux is $2.45\times10^{-10}$~ergs~cm$^{-2}$~s$^{-1}$ for this model. There was no significant improvement to the X-ray fits if we added additional components, such as a thermal accretion disk component or a Gaussian emission component in the 6--7 keV range. 

\subsection{Apache Point Observatory Optical and Near-Infrared Imaging}

We obtained optical and NIR photometry of \sj\ using SPIcam and NIC-FPS \citep{hearty2004} on the 3.5-m telescope at Apache Point Observatory, taking exposures in the B,V,R,I,J,H, and K filters. Weather conditions were clear with moderate, variable seeing (FWHM$\simeq$ 0.9$\arcsec$ -- 1.5$\arcsec$). We reduced the data using IRAF\footnote{IRAF is distributed by the National Optical Astronomy Observatories, which are operated by the Association of Universities for Research in Astronomy, Inc., under cooperative agreement with the National Science Foundation.} for standard calibration tasks (e.g., bias subtraction, sky subtraction, and flat-fielding) and aperture photometry. We calibrated the photometry in the optical using the in-field standards given in \citet{zurita2008}. For the NIR, we acquired separate standards from the ARNICA catalog \citep{hunt1998}. We averaged the results for multiple exposures for the target and standard star fields. The results are tabulated in Table~\ref{tab_oir}.  The uncertainties on the magnitudes are the propagated statistical errors for our exposures; because \sj\ is variable, the observed magnitudes fluctuate at levels that are larger than the statistical errors.

\section{Analysis} \label{sec_analysis}

\subsection{The Ultraviolet Spectrum}

\begin{figure}
\includegraphics[angle=90,scale=0.7]{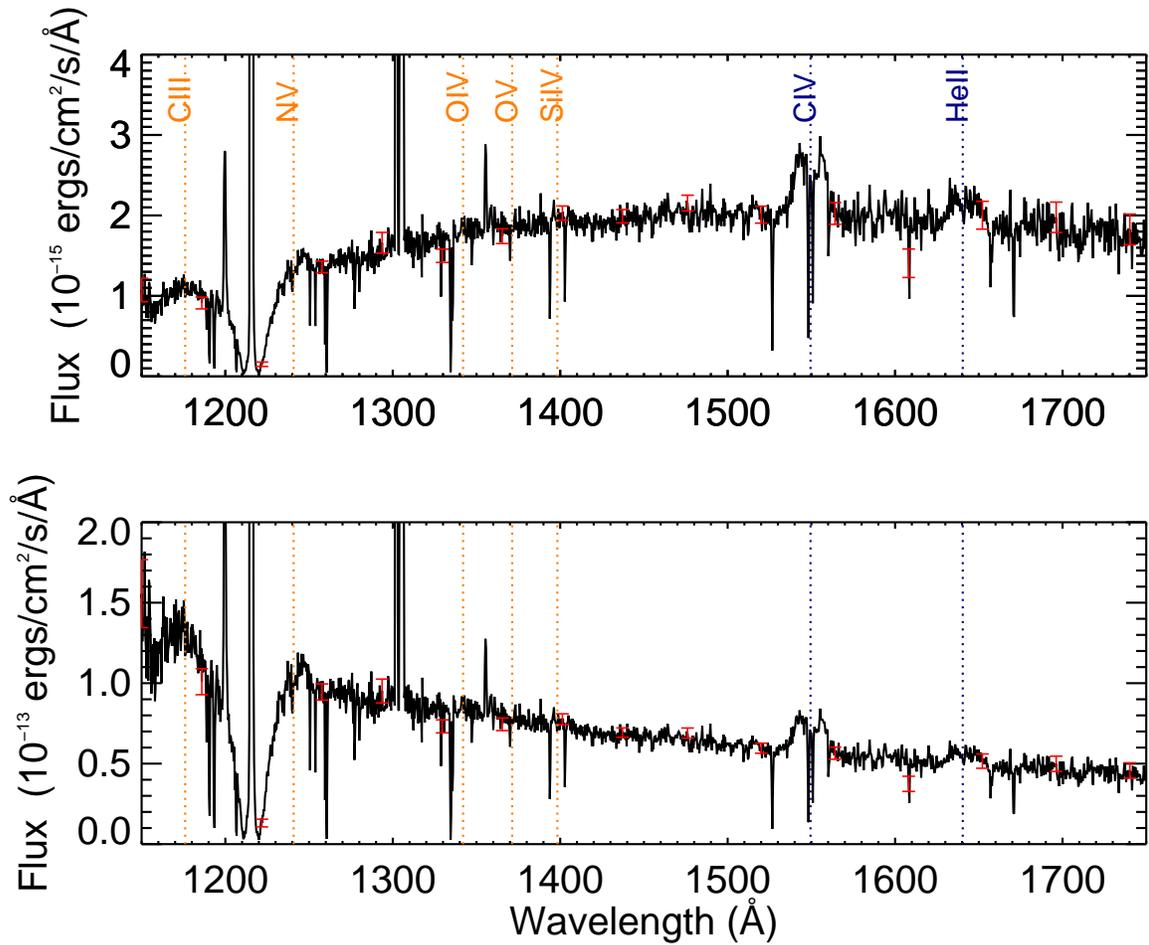}
\figcaption[combspec.ps]{The time-averaged COS FUV spectrum of \sj. The data have been binned to six resolution elements ($\simeq 0.36$~\AA) and typical statistical error bars on the binned data are shown in red. The upper panel shows the observed spectrum while the lower panel shows the spectrum after it has been dereddened assuming E(B--V) = 0.45. Prominent emission lines detected in the spectrum are labeled in blue. Emission lines often observed in the FUV spectra of other X-ray binaries but not detected unambiguously in \sj\ are labeled in orange. The narrow emission features at 1200~\AA, 1215~\AA\, 1304~\AA, and 1356~\AA\ are due to terrestrial airglow and the absorption lines are interstellar. \label{fig_cos}}
\end{figure}

The FUV spectrum of \sj\ is shown in Figure~\ref{fig_cos}. The observed spectrum has a FUV continuum flux of $2 \times 10^{-15}$~\flam\ at 1450~\AA. Spectral features include broad emission lines of \ion{C}{4}  $\lambda\lambda$1548,1551 and \ion{He}{2} $\lambda$1640 and interstellar absorption features \citep{morton1978}. In Figure~\ref{fig_cos}, we labeled the locations of other FUV lines that are often seen in the spectra of XRBs; however, only \ion{C}{4} and \ion{He}{2} are clearly detected in \sj. There is an enhancement in the spectrum around 1240~\AA\ that suggests the presence of the \ion{N}{5} $\lambda\lambda$1238,1242 doublet, but most of the line, if present, is lost in the broad interstellar Ly$\alpha$ absorption.   

\begin{figure}
\includegraphics[angle=90,scale=0.7]{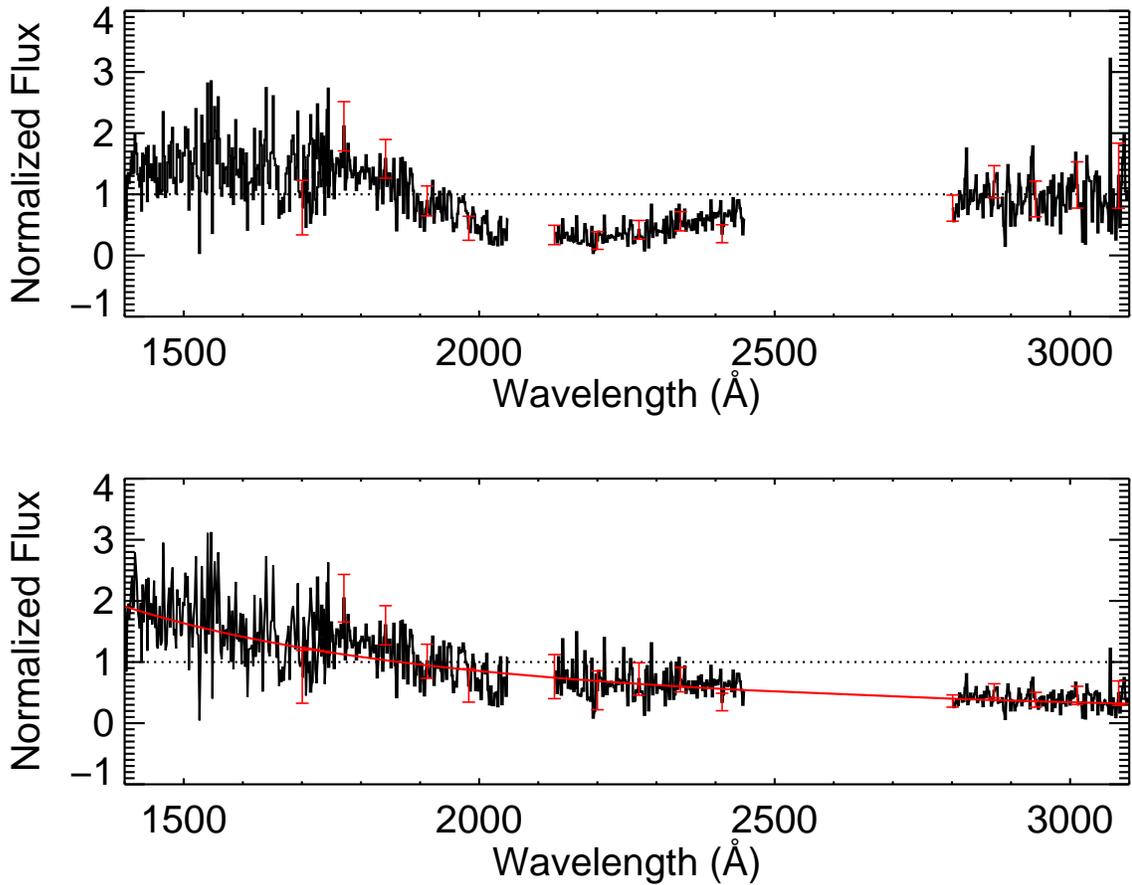}
\figcaption[dered.ps]{The COS NUV spectrum in the vicinity of the 2175~\AA\ interstellar absorption was fit to determine the reddening correction for the line of sight to \sj. The upper panel shows normalized G160M (from 1400--1750~\AA) and G230L spectra ($>1700$~\AA). The data have been binned to 1.6~\AA\ resolution. The lower panel shows the spectrum after it has been dereddened using the best-fit value of E(B--V) = 0.45. The best fit was found by minimizing the $\chi^{2}$ around a power-law fit to the dereddened spectrum, shown by the red line. \label{fig_red}}
\end{figure}

We dereddened the spectrum using E(B--V) = 0.45 (see \S~\ref{sec_ebv}). The dereddened spectrum shows a smooth continuum increasing to short wavelengths. The NUV spectrum extends the continuum to the red and shows no spectral features (see Figure~\ref{fig_red}). We fit single Gaussian profiles to the two emission lines in the dereddened spectrum using Specfit \citep{kriss1994}. The fit to \ion{C}{4} also included coincident interstellar lines to correct for their effects on the line flux. The \ion{C}{4} and \ion{He}{2} line fluxes are $5.5\pm0.8 \times 10^{-13}$~ergs~cm$^{-2}$~s$^{-1}$ and $2.6\pm0.6 \times 10^{-13}$~ergs~cm$^{-2}$~s$^{-1}$, respectively. Both lines have FWHM$\simeq 3500$~km~s$^{-1}$; the widths are not constrained to better than about 500~km~s$^{-1}$.  We do not have the signal to noise to determine if the emission lines are single or double-peaked. Using the observed flux density in the region of the \ion{N}{5} doublet and correcting for the interstellar absorption, we set an upper limit on the \ion{N}{5} line flux of $\lesssim 2.5 \times 10^{-14}$~ergs~cm$^{-2}$~s$^{-1}$ .

Using the COS time-tagged photon event list, we searched for variability in the continuum and emission lines during the observations. We identified regions of interest in the two-dimensional (wavelength vs. cross-dispersion position) corrected time-tag spectral images and summed the counts in each image for selected time bins, subtracting a background region of the same size but offset from the target spectrum in the cross-dispersion direction for each bin. We selected several line-free continuum regions and the \ion{C}{4} and \ion{He}{2} emission lines, examining the data in 50~sec and 200~sec time intervals.  We used different grating tilt settings during the observation, so we restricted our analysis to the time intervals when our chosen wavebands were fully on the detector. Because of the faintness of the target, we are insensitive to low-level variability: we could not detect fluctuations (at $>3\sigma$) below 10\% on 50~sec time scales. Above that limit, the COS data were steady over the course of the observations. There were no trends in the FUV light curves and all fluctuations were consistent with statistical noise. 

Finally, we examined the \ion{C}{4} line profile over time to see if we could identify radial velocity variations in the line over the orbital period. The orbital period of \sj\ is unknown: a 3.24~hr photometric periodicity has been observed and was interpreted as a superhump modulation that occurs in low mass ratio systems when a 3:1 orbital resonance is established within the accretion disk, causing the disk to precess and exhibit a periodic emission modulation \citep{zurita2008,haswell2001,whitehurst1988}. Using an empirical relation between the orbital and superhump periods determined by \citet{patterson2005}, Zurita et al.\ found an orbital period range for \sj\ between $3.18 < P_{orb} < 3.24$~hr. We examined spectra extracted in time subintervals from 100 to 800 sec. Unfortunately, the combination of low signal to noise in the individual spectra, the relatively low contrast between the line and continuum (the peak line flux is only 50\% above the continuum flux) and the contamination of the line center by the interstellar absorption precluded obtaining reliable velocity shifts from \ion{C}{4}. The \ion{He}{2} line was also too faint to yield significant results.

\subsection{Interstellar Absorption and Reddening} \label{sec_ebv}

There have been various estimates for the interstellar absorption and reddening along the line of sight to \sj. Fits to X-ray spectra in which hydrogen column density, $N_{H}$, was allowed to vary have resulted in values of $N_{H}$ = 1.7 -- 2.3$\times 10^{21}$~cm$^{-2}$ \citep{hiemstra2009,miller2006,morris2005}.
 \citet{cadolle2007} and \citet{durant2008} determined reddening values from the equivalent widths of the optical \ion{Na}{1} D lines using the prescription of \citet{munari1997} and converted these to $N_{H}$ after \citet{bohlin1978}. Using this method, \citet{cadolle2007} found E(B--V) = 0.34$\pm$0.04 and $N_{H} = 2.0 \times 10^{21}$~cm$^{-2}$, while \citet{durant2008} obtained E(B--V) = 0.42$\pm$0.02 and $N_{H} = 2.45 \times 10^{21}$~cm$^{-2}$. Durant et al.\ speculated that the absorption internal to the system may vary over time, given the difference in their results from those of Cadolle Bel et al. Although there is some variation in determinations of $N_{H}$, all of the derived values are comparable to the average total Galactic absorption along this line of sight \citep{morris2005,durant2009}.

We obtained values for both E(B--V) and $N_{H}$ from our UV spectra by fitting 2175~\AA\ dust feature and  the damping wings of the interstellar Ly$\alpha$ absorption line profile, respectively. To determine the reddening, we fit the 2175~\AA\ feature in the NUV. Figure~\ref{fig_red} shows the UV spectrum of \sj\ in the region of the 2175~\AA\ dust feature. The upper panel shows the observed spectrum normalized to a mean value of 1.0. Using the ccm\_unred task from the IDL Astronomy User's Library\footnote{\url{http://idlastro.gsfc.nasa.gov/homepage.html}} we dereddened the spectrum, stepping E(B--V) from 0.00 to 1.00 in steps of 0.01 to find the reddening value that minimized the $\chi^{2}$ deviation about a power law fit to the residual. $R_{V}$ was held fixed at 3.1. The best fit was E(B--V) = 0.45 with $\chi^{2}_{\nu} = 1.55$ (566 $dof$)for a 1400 -- 3100~\AA\ fit range. The reddening is consistent with the value found by \citet{durant2008}.  Following the lead of \citet{fitzpatrick1999}, who noted the large scatter about the mean in the amplitude of  the 2175~\AA\  feature in Galactic sight lines, we assume 20\% uncertainty on our derived reddening and adopt E(B--V) = $0.45\pm0.09$. 

\begin{figure}
\includegraphics[angle=90,scale=0.7]{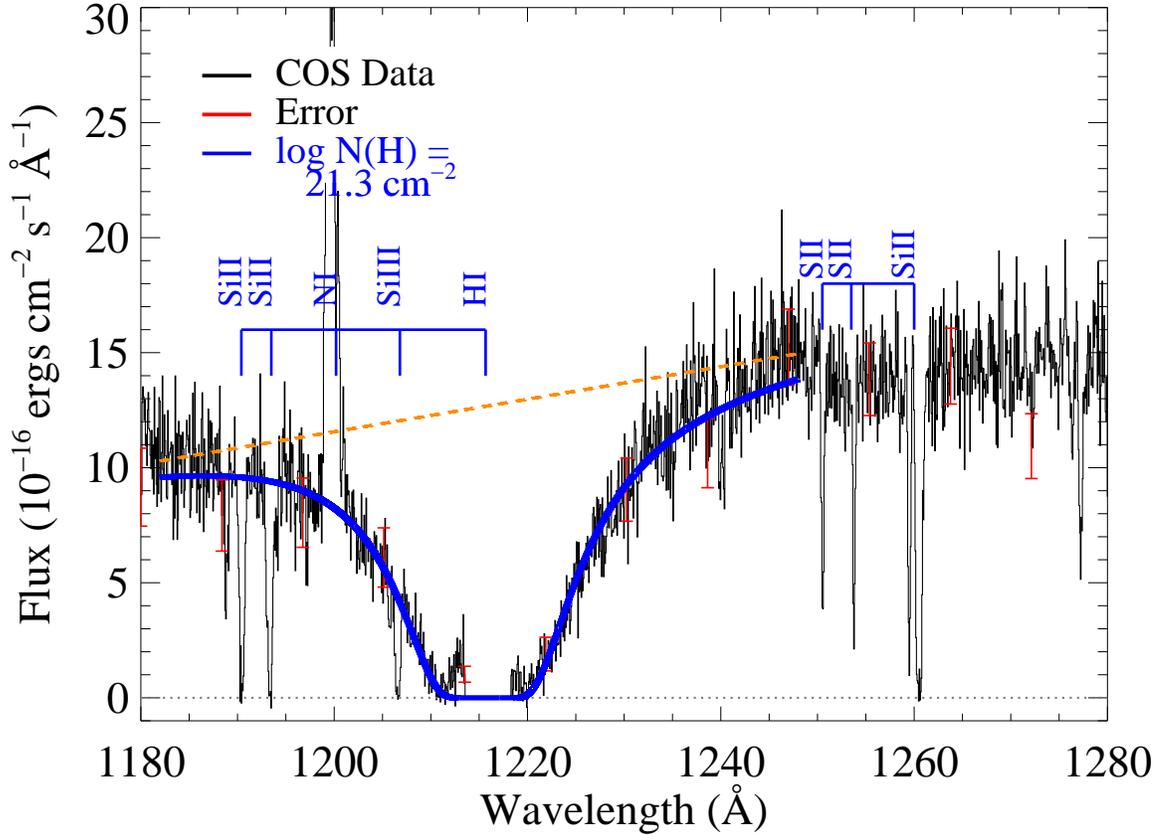}
\figcaption[lya.ps]{The Ly$\alpha$ absorption line in the COS spectrum is fit to determine the hydrogen column density on the line of sight to \sj. Interstellar absorption and terrestrial airglow lines that were masked out in the fit are labeled in blue. The center of the Ly$\alpha$ profile is contaminated by airglow and is masked out. The solid blue line shows the model fit to the damping wings of the line profile, while the dashed yellow line shows the continuum normalization across the line. \label{fig_lya}}
\end{figure}

Figure~\ref{fig_lya} shows the Ly$\alpha$ absorption line and the best fit to the line profile. We fit the interstellar \ion{H}{1} column density using a linear approximation of the observed (reddened) FUV continuum of \sj\ and a Voigt profile for the Ly$\alpha$ resonance line. First, the geocoronal airglow emission (1213.5 -- 1218.3~\AA) was removed from the core of the interstellar absorption trough.  A linear fit was then made across the Ly$\alpha$ region (1174~--~1270~\AA), anchored on the blue end by the average continuum flux from 1175~--~1186~\AA\ and on the red end by the average continuum flux from 1254~--~1258~\AA.  The Ly$\alpha$ absorber is characterized by a column density, $N_{H}$ (in units of cm$^{-2}$), a Doppler-$b$ parameter (in units of km s$^{-1}$), and a velocity offset relative to the rest wavelength of the Ly$\alpha$ transition (in units of km~s$^{-1}$ relative to $\lambda_{rest}$~=~1215.67~\AA).  The velocity was fit by eye to produce a profile that evenly filled in the red and blue edges of the Ly$\alpha$ absorption trough ($v_{Ly\alpha}$~=~$-$10 km s$^{-1}$).  The $b$-value was chosen to be typical of the local ISM (10 km s$^{-1}$; Redfield \& Linsky 2004), although the $b$-value does not have a large influence on the heavily damped line profile observed for \sj.  The interstellar column density was then varied until a best-fit value was found, with errors bars defined as values of $N_{H}$ that produce fits consistent with the 1-$\sigma$ photometric error bars in the line-core and wings.  This procedure yielded $N_{H} = 2.0\pm0.3 \times10^{21}$ cm$^{-2}$.

If we use the reddening to infer $N_{H}$ from the relation of \citet{bohlin1978}, we obtain a value larger than the one directly found from the Ly$\alpha$ line profile fitting, although they agree within the uncertainties on $N_{H}$ and E(B--V). Relating gas and dust absorption is subject to a number of uncertainties that could cause the slight discrepancy: we do not know the value of $R_{V}$ for this sight line and there is considerable scatter in the relationships between reddening and the size of 2175~\AA\  feature (as noted above) and between gas and dust for individual sight lines  ($\sim$30\% scatter about the Bohlin et al.\ relation). The gas along the line of sight to \sj\ may also be more metal rich than the average Galactic value, resulting in a larger reddening and a higher N$_{H}$ found from the X-ray fitting compared to the fit to the Ly$\alpha$ line profile (whereas the Ly$\alpha$ interstellar line directly traces the neutral hydrogen along the line of sight, the $N_{H}$ from the X-ray model fits includes absorption from helium and metals). Ultimately, the various methods for determining  $N_{H}$ and E(B--V) for the sight line to \sj\ have proven satisfactorily  consistent within the uncertainties.

\subsection{Accretion Disk Model Fits to the UV Spectrum} \label{sec_diskmod}

If we assume that the UV emission is dominated by the accretion disk, we can use disk model fits to the dereddened spectrum to determine black hole masses and accretion rates as a function of the distance to \sj.  Most parameters are unknown or poorly constrained for this system, however, so our fits will be more instructive in placing limits on interesting values than in providing precise determinations. We fit the dereddened UV continuum with a grid of steady-state accretion disk model spectra constructed from summed, area-weighted blackbody spectra. The blackbody spectrum for each disk annulus was chosen to correspond to the theoretical, steady-state thin accretion disk temperature for that annulus. The models did not include disk irradiation. 

Following \citet{cheng1992}, we can describe the observed flux density from a thin, steady-state, viscously-heated accretion disk as:

\begin{equation}
f_{\nu} = f_{0} \frac{\cos i}{d^{2}} (m\dot{m})^{2/3} \nu_{15}^{1/3} \int_{x_{in}(\nu)}^{x_{out}(\nu)} \frac{x^{5/3}dx}{e^{x} - 1}
\end{equation}

where $f_{0} \simeq 2.9 \times 10^{-26}$~ergs~cm$^{2}$~s$^{-1}$~Hz$^{-1}$, $d$ is the distance in $kpc$, $i$ is the binary inclination, m (= M$_{BH}$ / \msun) is the mass of the black hole, $\dot{m}$  is the mass accretion rate in units of 10$^{-9}$~\msun~yr$^{-1}$, $\nu_{15} = \nu / 10^{15}$ Hz, and $x = h \nu / k T(r)$ where $T(r)$ denotes the blackbody temperature (K) and radius (in units of 10$^{11}$ cm) of each annulus.  

The disk temperature at each annulus is a function of M$_{BH}$, $\dot{m}$, and the radius. The overall temperature distribution in the disk depends on the inner and outer disk radii adopted. For the inner disk radius, r$_{in}$, we examined models between two extreme values: r$_{in} = 1.23 r_{g}$ and 500~r$_{g}$, where r$_{g}$ is the gravitational radius for a given black hole mass. The values represent differing predictions of the inner radius of the thin disk as extending to the Innermost Stable Circular Orbit (ISCO) for a maximally-rotating Kerr black hole or truncated at larger radii under the ADAF model paradigm \citep{reynolds2010, zhang2010}. We set the outer disk radius, $r_{out}$, to 60\% of the Roche lobe radius. Following Zurita et al., we fixed $P_{orb}$ = 3.23~hr and the donor star mass to $M_{2} = 0.3$~\msun, which we used in conjunction with the black hole mass to set the binary geometry and $r_{out}$.\footnote{\citet{shaw2013} have claimed a 402~d modulation in the X-ray light curves of \sj. If this is attributed to a disk precession period, the inferred mass ratio for the system may be very low, $q \sim 0.002$, and M$_{2}$ may be much smaller than 0.3~\msun. This has a negligible effect on our models here, however, as the UV emission is only weakly dependent on the size of the outer accretion disk.} Because the UV flux is dominated by emission from the inner disk annuli, the exact value of the $r_{out}$ does not affect our fit results.

%The final value depends on the mass ratio and changed for each black hole mass we tested (ranging from $5\times10^{10}$ to $\simeq10^{11}$~cm), but because the UV flux is dominated by emission from inner disk annuli the exact value of r$_{out}$ does not affect the fit results significantly.  

For these model parameters, the accretion disk model spectra for \sj\ follow a $\nu^{1/3}$ power law profile in the FUV spectral region. We can therefore use the normalization of the disk models to the observed FUV flux to constrain the black hole mass as a function of accretion rate, distance, and inclination. For the continuum region centered on 1470~\AA\ (1425--1520~\AA; $\nu = 2.04\times10^{15}$), we measure an average flux density of $f_{\lambda} = 6.48\times10^{-14}$~ergs~cm$^{2}$~s$^{-1}$~\AA$^{-1}$. Converting to the frequency domain and substituting into equation 6 of  Cheng et al., we can express the mass of the black hole in \sj\ as:

\begin{equation}
\frac{M_{BH}}{M_{\sun}} = 0.546 [\cos i]^{-3/2} \left[\frac{d}{1 kpc}\right]^{3} \left[\frac{\dot{m}}{10^{-9} M_{\sun} yr^{-1}}\right]^{-1}
\end{equation}

The binary inclination, distance, and mass accretion rate for \sj\ are not known. The system does not show eclipses, which restricts inclinations to $i \lesssim 80^{\circ}$.  \citet{reis2009} found $i = 55^{\circ +2}_{ -7}$ based on model fits to X-ray reflection features. We adopt this as our default inclination but also examine how varying the inclination affects the other model parameters. (Note that the reflection fits assume that the disk extended to the ISCO at the time of the observations, which may not have been the case; see \S~\ref{sec_discussion}.) 

\sj\ has remained in the low/hard state throughout its outburst. \citet{maccarone2003} found that XRBs transition from the soft to the hard state between 1--4\% of the Eddington luminosity.  The transition from hard state to soft can occur over a somewhat larger range of luminosities due to hysteresis effects \citep{maccarone2003b}.  
%In 2009, \sj\ experienced a ``failed transition" in which the X-ray spectrum softened, moving from the low/hard to an intermediate state, though it never fully entered the soft state \citep{soleri2013}.
Here, we adopt an upper limit on the mass accretion rate of $\dot{m} \leq 0.05 \dot{m}_{Edd}$, where $\dot{m}_{Edd} = L_{Edd} / (0.1 c^{2})$, assuming a radiative efficiency of 10\%. For the units in equation (2), this corresponds to $\dot{m} \leq 1.097 \times m$. Since our data were taken at a X-ray luminosity several times smaller than the peak value (when the system remained in the low state), this upper limit is conservative.

Based on these parameter assumptions, we have generated upper limits on the distance to \sj\ for given black hole masses based on fits to the dereddened UV continuum. These are given in Table~\ref{tab_disk}. For the upper limits, we also assumed that the disk radius extends to the ISCO to maximize the potential disk emission. The distance could be higher for a lower inclination disk so we also present the upper limit on the distance for the extreme case of a face-on disk with $i=0^{\circ}$.  

\begin{figure}
\includegraphics[angle=90,scale=0.7]{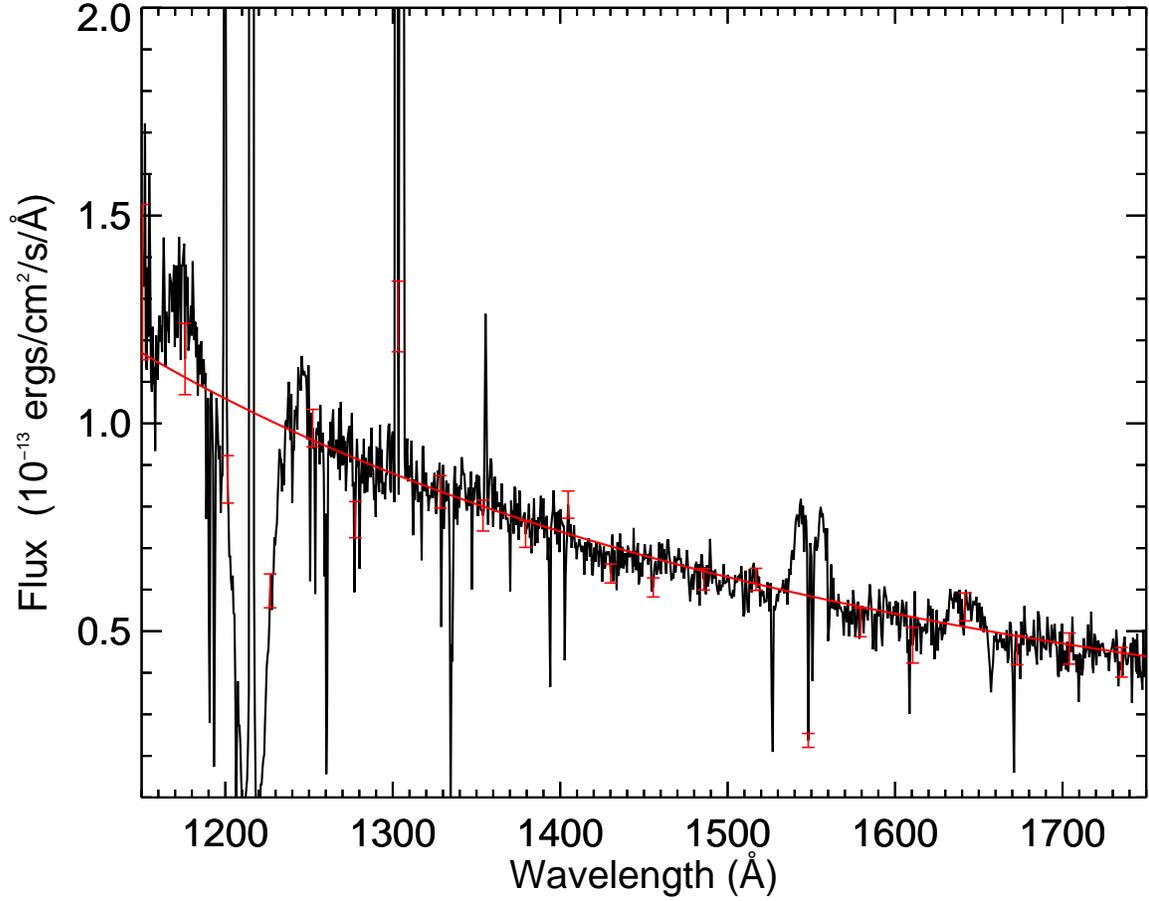}
\figcaption[diskmod.ps]{The dereddened spectrum of \sj\ compared to a steady-state accretion disk model spectrum. The model has $M_{BH} = 12.0$~\msun, d = 5.1~kpc, and an inner disk radius of $r_{in} = 1.23 r_{g}$. The inclination is $i = 55^{\circ}$. The spectrum has been binned to 0.5~\AA\ spectral resolution. The model has not been formally fit to the spectrum but is overlaid by eye. \label{fig_diskmod}}
\end{figure}

Figure~\ref{fig_diskmod} shows the 12~\msun\ disk model compared to the data. The accretion disk models with viscous heating and no irradiation provide qualitatively good fits to the shape of the FUV continuum.  The observed spectrum may be somewhat more blue than the model at the shortest FUV wavelengths, but we don't make a quantitative conclusion on this point, because a small error in the adopted dereddening and/or the presence of unidentified emission features (such as \ion{C}{3} 1175~\AA\ and \ion{N}{5} 1238, 1242~\AA) could bias the continuum shape in this region. We are unable to distinguish between an accretion disk extending to the ISCO or one truncated at larger radii from the UV data alone (although see \S~\ref{sec_discussion} for comparisons with the broadband SED). In either scenario, the UV is on the $\nu^{1/3}$ power law portion of the disk SED, with peak emission occurring in the EUV for the truncated disk or in soft X-rays for the ISCO disk. The two radii give different results when fit to the data, however, since the ISCO model generates more UV flux: e.g., for a 9~\msun\ black hole at a 6~kpc distance, the inferred mass accretion rate must be 25\% higher for the truncated disk compared to the ISCO radius to match the observed UV flux. 

\subsection{The Spectral Energy Distribution} \label{sec_sed}

\begin{figure}
\includegraphics[angle=90,scale=0.7]{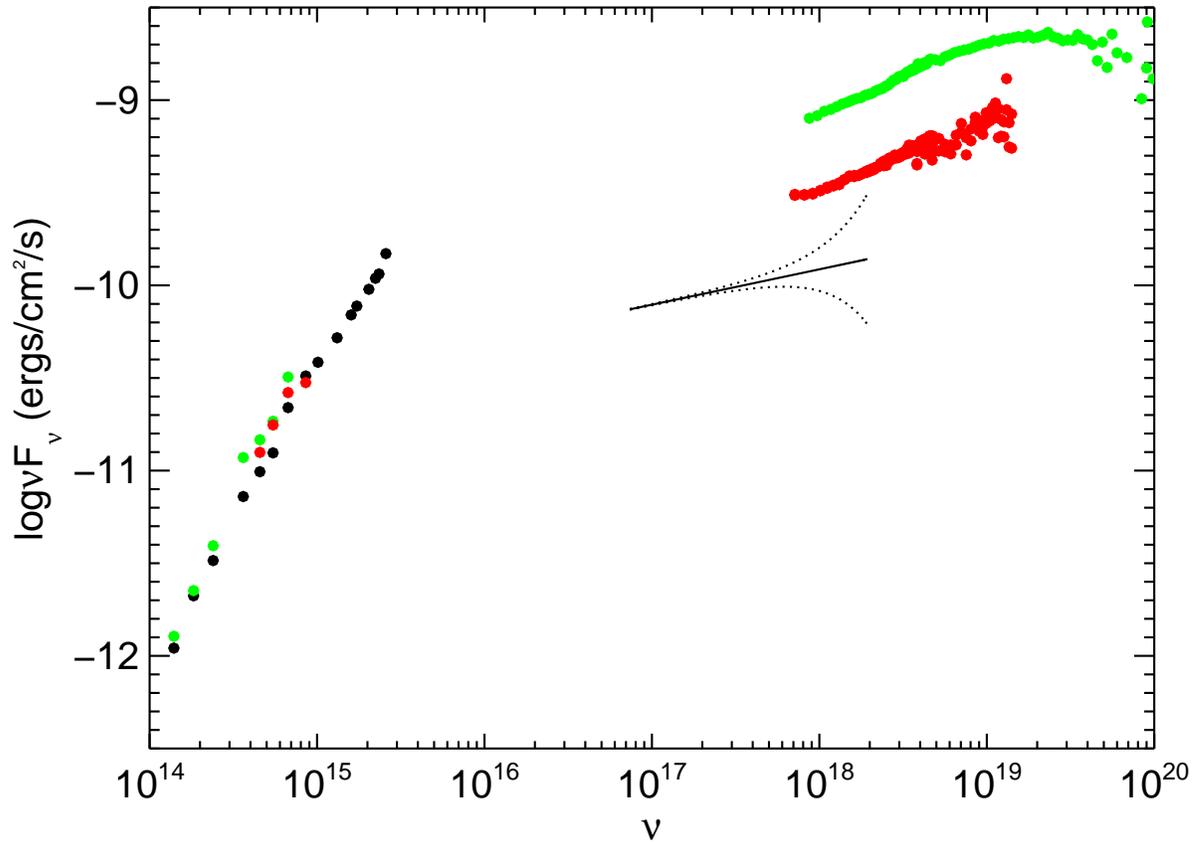}
\figcaption[sed.ps]{The broadband spectral energy distribution for the 2012 observations of \sj\ is shown in black. The NIR, optical, and UV data have been dereddened assuming E(B--V) = 0.45. The X-ray data show the power-law fit  to the Swift XRT observations with $N_{H}$ fixed at $2.0\times10^{21}$~cm$^{-2}$. The dashed lines show the uncertainties on the X-ray data. Shown in green and red, respectively, are the data from \citet{cadolle2007} and \citet{durant2009}. \label{fig_sed}}
\end{figure}

In Figure~\ref{fig_sed} we plot the broadband spectral energy distribution (SED) for \sj\ at the time of our observations.  Table~\ref{tab_sed} gives the data we used for the SED for future reference. The NIR/optical/UV data were dereddened assuming E(B--V)=0.45. For the X-ray data, we plot the power law component of the model fit to the data and its uncertainties.  In the figure, we also plot previous SEDs acquired in 2005 (three months after outburst start) and in 2007, with the optical/NIR points dereddened using our E(B--V) value for a consistent comparison \citep{cadolle2007,durant2009}. 

A comparison of the X-ray SED from the three epochs shows that the X-ray fluxes have declined. The system has also faded in the optical/NIR. The NIR fluxes we observed are $\simeq$90\% and the optical fluxes are $\simeq$65\% of those observed in 2005. The 2012 data are also fainter than the 2007 observations, with fluxes that are $\simeq$75\% of the 2007 data in the optical. The shape of the optical/NIR SED has not changed significantly as it has faded, however: all three epochs show approximately power law shapes, although a single power law cannot fit all of the optical/NIR data in any epoch. The new FUV observations show that the power law shape extends to $\simeq$1150~\AA\ at least: the spectral break between the X-rays and lower energy data occurs at higher energies than the FUV. 

\section{Discussion} \label{sec_discussion}

\subsection{The UV Spectrum}

In this manuscript, we have presented the first UV spectrum of \sj, accompanied by contemporaneous X-ray, optical, and NIR observations. The dereddened UV spectrum is characterized by a continuum that smoothly increases to the blue and broad line emission from \ion{C}{4}, \ion{He}{2}, and possibly \ion{N}{5}. There are now high quality FUV observations of three black hole XRBs in outburst: \sj, XTE J1118+480, and XTE J1859+226 \citep{haswell2002}. The emission line spectra are different in each case. In XTE J1859+226, the UV spectrum has strong lines of \ion{C}{3}, \ion{C}{4}, \ion{N}{5}, \ion{O}{5}, \ion{He}{2}, and \ion{Si}{4}. That spectrum is richer in emission lines than \sj, and the \ion{N}{5} line flux is nearly equal to that of \ion{C}{4}, whereas we do not confidently detect \ion{N}{5} here. As noted by Haswell et al.\ the \ion{N}{5} line can be suppressed relative to \ion{C}{4} by either photoionization by a harder X-ray ionizing spectrum or by relatively metal-poor abundances. 

On the other hand, XTE J1118+480 shows the opposite effect, with much stronger \ion{N}{5} emission and no \ion{C}{4}, an effect that has also been observed in \vmon\ \citep{froning2011} and is attributed to CNO processing in the accreted material.  If \sj\ had followed the evolutionary path proposed by Haswell et al.\ for XTE J1118+480 and A0620-00, wherein mass transfer was initiated after the donor star had already begun to move off the main sequence, their model would predict a current C/N abundance ratio for \sj\ of $\log (C/N) \leq -2.3$, lower than in XTE J1118+480, where \ion{C}{4} is undetectable (Haswell et al.\ Fig. 3; see also their discussion of why the low \ion{C}{5} to \ion{N}{5} ratio is unlikely to be caused by photoionization effects). Instead, the UV line spectrum of \sj\ suggests little CNO processing, which is consistent with a system that initiated mass transfer at a shorter orbital period when the lower-mass donor star was still on the main sequence. More generally, the comparison of the three black hole XRB spectra in the UV shows that the complicated interplay of evolutionary history, metallicity, and photoionization leads to diverse spectroscopic signatures; observations of more systems will be valuable to constrain the physical processes underlying the line emission.

%Four halo sources: XTE J1118+480, Swift J1357-0933, Maxi J1659-152, and \sj. They share high Galactic latitude, short orbital period, and (for all but Maxi) outbursts that never left the low/hard state. Not sure if there is significance here. Gonzalez 2008 favor a disk origin plus natal kick for 1118+480 because of metal-rich abundances and Kuulkers et al. 2013 also favor the kick idea and say lower mass companions are needed for short period systems and are Äalso easier to kick further. They don't seem to consider in situ birth, though: their alternative is a globular cluster origin.

\sj\ was stable during our observations to $<$10\%, showing neither secular variability nor any evidence of the 3.24~hr  orbital modulation seen in the optical light curves \citep{zurita2008}. The orbital modulation may have been below our detection threshold, however. Our longest continuous light curve in a single waveband was 5538~sec, or about half of the 3.24~hr modulation. The Zurita et al.\ R-band light curves typically showed a 0.1~mag or smaller variation over those time scales, which is at the level of our sensitivity, assuming a similar amplitude in the FUV. Our NUV spectra show hints of variability on the $\sim$8\% level but the 1400~sec baseline in each spectrum is too short to determine if the orbital modulation is present. It is also unclear whether the modulation, believed to be due to superhumps, will still be present in \sj\ several years after the initial outburst, although superhumps were seen in the light curve of XTE J1118+480 near quiescence \citep{zurita2002}. The expected amplitude of the superhumps will likely be lower in the UV than in the optical, as the effect occurs in the cooler, outer disk. \citet{haswell2001} attributed superhumps in XRBs as an effect of changing disk area rather than changes in viscous dissipation in the disk, which might be more visible in the UV; however, this effect was in play for XRB disks dominated by irradiation, which may not be the case here (see below).

In \S~\ref{sec_diskmod}, we fit the dereddened UV continuum spectrum with models of emission from a steady-state thin accretion disk. The models provide qualitatively acceptable fits to the spectral shape. The assumption that the UV emission is dominated by the thin disk is consistent with observations at other wavelengths and with system models.  \citet{reynolds2010} showed that a power law with a slope consistent with optically thin emission from a jet extended from the X-ray regime to longer wavelengths would have a negligible contribution to the optical/NIR flux. Their notional power law component falls roughly an order of magnitude below our dereddened FUV fluxes. Similarly, an ADAF model fit to \sj\ also predicts UV emission dominated by the thin disk \citep{zhang2010}. 
%Optical/X-ray cross-correlation analyses have also been interpreted as evidence that the optical waveband is dominated by emission from the disk \citep{durant2008,hynes2009}. 
Finally, \citet{soleri2010} found that the jet is fainter than in \sj\ than in other hard state XRBs based on its radio/X-ray flux ratio compared to the standard empirical relation. They too attribute the NIR/optical emission to the accretion disk. 

Our disk models fit the UV continuum well with vicious dissipation in the disk as the only emission source. Disk models including irradiation predict a flattening in the SED in the FUV that is inconsistent with our observed spectrum (see, e.g., Figures 7 and 9 of Chiang et al. 2010). This places \sj\ in the category of ``UV-hard" sources as defined by \citet{hynes2005}, who compiled the UV/optical SEDs of several black hole XRBs in outburst and found that the systems either showed rising fluxes in the FUV (the UV-hard sources) or a peak in the NUV with a drop off in flux to shorter wavelengths (the ``UV-soft") systems. The UV-hard sources were fit with viscously-heated accretion disks while the UV-soft sources were dominated by emission from reprocessing of X-ray irradiation in the disk. \citep[The UV-hard spectra can also be fit by irradiation models if the irradiating source is vertically extended above the disk plane, but the characteristic spectrum in that case remains different from the UV-soft profile;][]{hynes2005,dubus1999}. One system, XTE J1859+226, has shown a transition during the outburst decline from being best fit with irradiation-dominated model spectra to models dominated by viscous heating \citep{hynes2002}. 
%\citet{chiang2010} fit irradiated disk models to optical and X-ray observations of \sj\ which predict a flattening of the SED in the FUV, contrary to the smooth extension of the power law profile we observe at those wavelengths. (Chiang et al.\ do not, in fact, prefer the irradiation models for their late observations, two years after the initial outburst, but their fits are illustrative of the difference between predictions of irradiated disk models and observed FUV fluxes in \sj.) 
The X-ray/optical cross-correlation function in \sj\ also evolved from a clear signature of thermal reprocessing near outburst peak to an unusual profile at later times in which the optical and X-ray emission are anti-correlated, with the optical leading the X-rays \citep{hynes2009,durant2008}. A similar cross-correlation function was seen in XTE J1118+480, another XRB that remained in the low/hard state during its outburst \citep{kanbach2001}. These correlations indicate a link between optical and X-ray emission, but not one that is explained by simple reprocessing of X-rays in the outer accretion disk.

Because the system parameters for \sj\ are not well constrained, we could not use our disk model fits to directly determine the mass accretion rate in the disk. However, by setting upper limits on the accretion rate as a fraction of the Eddington luminosity based on the fact that the system did not transition from the hard state to the soft state during the outburst, we were able to place limits on the mass of the black hole and the distance to the system based on the dereddened FUV continuum fluxes. We presented upper limits on the distance to \sj\ for a range of black hole masses. These values are consistent with those found by \citet[][note that they present the minimum black hole mass for a given distance while we give the maximum distance for a given mass]{cadolle2007}.
%The X-ray and UV data trace two different accretion regions --- the hot, inner flow and the thin disk, respectively --- but the luminosities of the two wavebands make comparable predictions for the black hole mass-distance relation   

We are in closest agreement with the X-ray predictions for our models in which the disk is face-on rather than inclined. The Cadolle Bel et al.\ observations were conducted near outburst peak while ours were taken during the ongoing state in which \sj\ has declined from its peak level but has not returned to quiescence. Accordingly, we expect that the disk accretion rate is lower now than in 2005, which is why our low inclination models (which produce more observable UV flux) are in closer agreement with the earlier black hole mass-distance relation. For a higher inclination, our more recent observations place tighter limits on the distance for a given black hole mass. Our limits are also conservative, since the mass accretion rate should be well below the 5\% of Eddington transition luminosity seven years after the initial outburst. The hysteresis effect that can boost the transition luminosity for the hard to soft transition relative to the soft to hard one is a phenomenon of early outburst and will be seen near the peak rather than during the outburst decline \citep{maccarone2003b}. Moreover, \citet{soleri2013} found that the X-ray spectrum of \sj\ began to soften, but did not fully transition to the soft state, at an X-ray flux that was $\simeq$5 times larger than our observed fluxes. 

Although we cannot directly determine the black hole mass from our observations, our black hole mass/distance limits can be compared to current estimates of the distance to \sj, which lie roughly in the d = 1 -- 10~kpc range. \citet{zurita2008} used the non-detection of \sj\ in optical catalogs prior to outburst to set the quiescent magnitude V$_{quies} > 21$ and a distance $d >1$~kpc for a M2 main sequence donor star. Their comparison of the outburst luminosities to various empirical relations for XRBs gave a distance range of $\simeq$2.5--8~kpc. The interstellar hydrogen absorption column density in the spectrum, $N_{H}$, of \sj\ is comparable to the total Galactic column density in this direction, also suggesting a distance $d \sim$6--7~kpc \citep{morris2005,cadolle2007,durant2009}.  A 6~kpc distance would correspond to a black hole mass $\geq$10.4~\msun.

We compared our disk model results to the system parameters adopted by \citet{zurita2008}: $M_{BH} = 12.0$~\msun\ and $d=5.5 kpc$.  If we adopt those values, we must set $\dot{m} \simeq 2 \times 10^{-8}$~\msun / $yr$, or $0.08 \dot{m}_{Edd}$, to match the UV flux (for an $i = 55^{\circ}$ disk). Some of the ADAF models of XRBs have predicted that the accretion rate could be as high as $\dot{m} = 0.08 - 0.1 \dot{m}_{Edd}$ in the hard state \citep{esin1997, zhang2010}. If the accretion rate is $<$8\% Eddington, the Zurita et al.\ parameters would require an adjustment to a higher black hole mass or a lower distance to be consistent with the UV fluxes. Once \sj\ has returned to quiescence, there may be opportunities to place better constraints on the dynamical parameters of the binary system if the donor star can be observed. In that event, it will be of interest to revisit disk model fits to the UV spectrum to place tighter limits on the mass accretion rate through the thin disk during the outburst.
% and perhaps distinguish between models with the thin disk extending to the ISCO and one truncated at larger radii. 

\subsection{The Broadband SED}

\sj\ has faded at all wavelengths since 2007, when previous broadband data were acquired \citep{durant2009,soleri2010}. The X-ray spectrum has dropped in luminosity by $\simeq$60\%. 
%and the power law spectral index has increased slightly (although note we did not fit the same spectral band as in  the previous observations). 
The optical/NIR data continue the slow decline seen since the 2005 outburst \citep{shaw2013}. The shape of the SED has not changed dramatically, although the flux decline has been larger in the optical than in the NIR. The thin disk component fit to the UV also fits the optical fluxes reasonably well. There is a slight excess in the NIR compared to the model. Some of the excess NIR flux can be recovered by increasing the outer disk radius above the $0.6 R_{L_{1}}$ value we used, but even disks extending to 90\% of the Roche lobe radius don't recover all of the NIR emission. The excess may indicate an additional contribution from the jet, from the hot flow component of an ADAF, or from synchrotron emission by non-thermal electrons in the hot corona \citep{zhang2010,veledina2013}. 

\begin{figure}
\includegraphics[angle=90,scale=0.7]{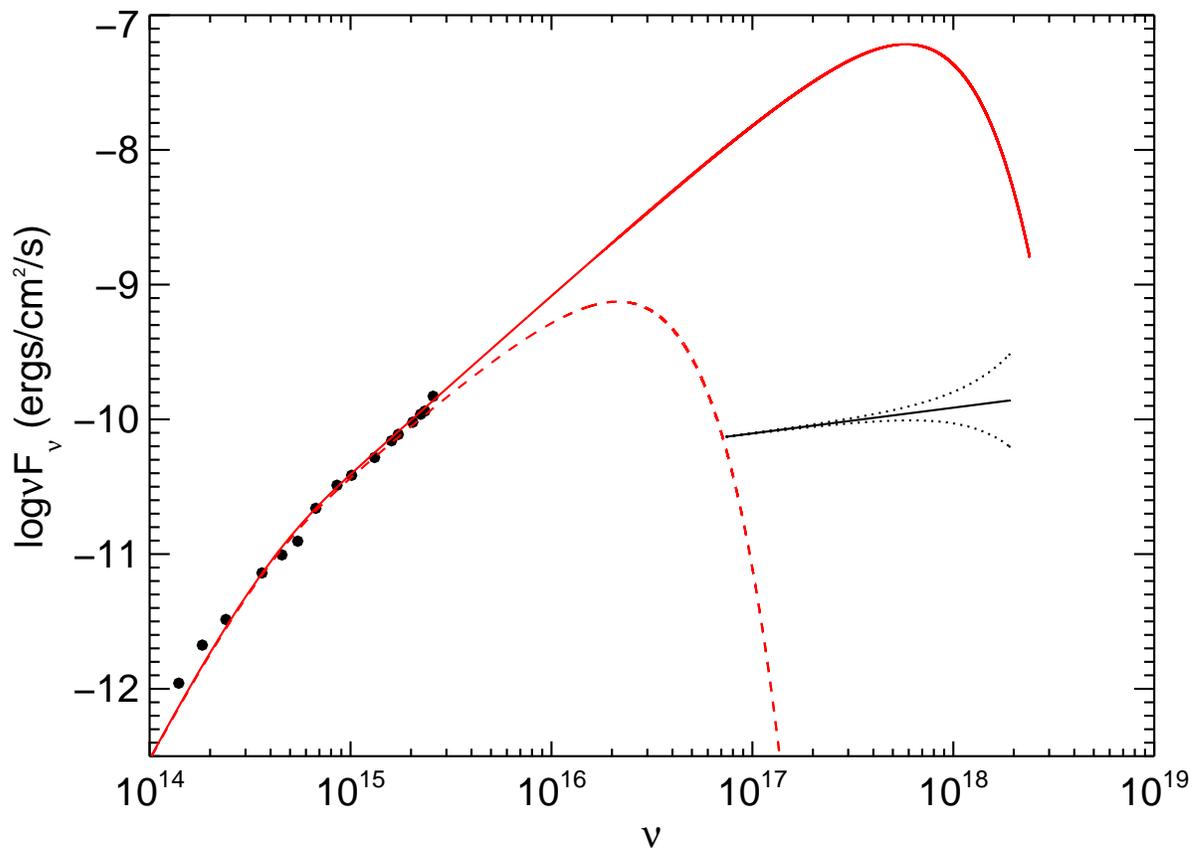}
\figcaption[sedcmp.ps]{Steady state accretion disk models are compared to the observed SED of \sj. The solid red line shows the 12.0~\msun\ model shown in Figure~\ref{fig_diskmod} extended over the broadband SED. The dashed red line shows a model with the same parameters, except that the inner disk radius has been truncated at $r_{in} = 100 r_{g}$. \label{fig_sedcmp}}
\end{figure}

As noted in \S~\ref{sec_diskmod}, we could not distinguish between accretion disks whose inner radii extend to the ISCO versus those that are truncated at a larger radius from the UV data alone.  However, if we compare the disk models to the broadband SED, we find a discrepancy between the observed UV and X-ray fluxes if the accretion disk extends to the ISCO. This is illustrated in Figure~\ref{fig_sedcmp}, where we plot our 12~\msun\ model on the SED. The model that fits the UV fluxes well exceeds the X-ray component by two orders of magnitude. 

There are two ways to resolve this discrepancy. First, if the UV is not dominated by the steady-state thin disk, a disk that extends to the ISCO can be present. The models of \citet{miller2006} and \citet{reynolds2010}, for example, include a thermal component from a thin disk extending to the ISCO.  Their disk components (assuming a 10~\msun\ black hole and 8.5~kpc distance) will not match the UV: they fall a factor of 5 below our 2011 observed fluxes, which will moreover be fainter than the 2006/2007 data fit by these groups. In this event, the NIR/optical/UV data would need a substantial contribution from an additional component with a power-law shape. In some XRBs, the UV emission is dominated by a single temperature blackbody component with a small emitting area, attributed to the hotspot where the in falling accretion stream impacts the outer disk or to emission at the transition radius between the thin disk and the ADAF \citep{froning2011, mcclintock1995,hynes2012}. These systems were in quiescence, however, not the active state observed in \sj, and a single temperature blackbody does not fit the slope of the UV data.  Given that \sj\ is in outburst, has a UV SED consistent with steady-state accretion disk models, and, as discussed above, other emission components (such as the jet or the X-ray hot flow) are unlikely to contribute substantially at these wavelengths, we believe that the UV emission is dominated by the thin disk.

Second, the disk could be truncated at a larger radius than the ISCO.  In Figure~\ref{fig_sedcmp}, we show that a disk model whose inner radius is truncated at $r_{in} = 100 r_{g}$ fits the optical/UV data without exceeding the X-ray flux. The inner radius is larger than the $r_{in} = 10 - 50 r_{g}$ radii cited by \citet{reynolds2010} but not as large as the $250 r_{s}$ ($500 r_{g}$) radius adopted by \citet{zhang2010} for their ADAF model fits. 
%If the black hole mass is substantially larger than the 5--15~\msun\ range we focused on in our fits, the disk could extend closer to the ISCO. A 50~\msun\ black hole located at $\simeq$13~kpc would fit the UV observations. Some disk truncation is still required ($r_{in} = 60 r_{g}$ in this example) to keep the X-ray fluxes at or below the observed level. 
The low luminosity in the X-ray band relative to the UV underscores the importance of understanding the energy balance in the low state of XRBs near the black hole: significant mass loss from the system and/or enormous energy loss via advection is required to reconcile the observational discrepancy.

\citet{chiang2010} also noted that the X-ray data in \sj\ are under-luminous relative to optical/UV fluxes, with accretion disk model fits to the X-rays under predicting the longer wavelength data. The deviation between the extrapolated X-ray model fits and the optical/UV data increased with time further from outburst peak. They were able to fit the SED with a truncated, irradiated disk to fit the optical/UV and invoked a separate soft component to fit the X-rays. They did not favor that model because the irradiated area of the disk was implausibly small and instead attributed the optical/UV data to synchrotron radiation from the jet. As we noted above, however, the FUV data are inconsistent with the irradiated disk models. Given that the jet in \sj\ is under-luminous compared to other hard state XRBs, and the radio to optical spectral indices are inconsistent with the compact steady jet extending to the NIR/optical waveband, the evidence argues against jet-dominated emission in the optical/UV  \citep{soleri2010}.

\section{Conclusions} \label{sec_conclusions}

We have obtained the first ultraviolet spectra of the black hole X-ray binary, \sj, using the Cosmic Origins Spectrograph on the Hubble Space Telescope. The dereddened UV spectrum has a blue continuum and broad (FHWM$\simeq$3500~km~s$^{-1}$) emission lines of \ion{C}{4}, \ion{He}{2}, and possibly  \ion{N}{5}. The UV count rate was stable to $<$10\% during the observations. We fit the 2175~\AA\ interstellar feature to determine an estimate of the interstellar reddening as E(B--V) = 0.45$\pm$0.09. By fitting the interstellar Ly$\alpha$ line profile, we also derived the neutral hydrogen column density along the line of sight to \sj\ as N$_{H} = 2.0\pm0.3 \times 10^{21}$~cm$^{-2}$. 

We fit models of a steady state accretion disk to the dereddened UV continuum. A viscously heated disk with no reprocessing of X-ray irradiation included provided a good fit to the shape of the spectrum. The binary system parameters are not well constrained, but by assuming that the mass accretion rate of \sj\ (which has never transitioned out of the low/hard state during the outburst) is below 5\% of the Eddington rate, we placed limits on the distance for a range of black hole mass values. Based on estimates of the distance to the source, a black hole mass well above 3.0~\msun\ is favored, although the lower mass values are not fully ruled out. 

We presented a broadband NIR/optical/UV/X-ray spectral energy distribution of \sj. The source has faded at all wavelengths since previous broadband observations in 2007, and the X-ray power law spectral index has steepened. The UV component of the SED extends the power law seen in the optical to shorter wavelengths with no evidence of a spectral break. However, the thin disk component required to fit the UV fluxes will exceed the observed X-ray data unless some truncation of the inner disk beyond the ISCO is adopted and large mass loss from the system or energy loss via advection into the black hole takes place.

\acknowledgments
Thanks to M. Cadolle Bel and M. Durant for sharing their spectral energy distribution data. R.I.H. acknowledges support from the National Science Foundation under Grant No. AST-0908789. Based on observations made with the NASA/ESA Hubble Space Telescope, which is operated by the Association of Universities for Research in Astronomy, Inc., under NASA contract NAS 5-26555. These observations are associated with program 12039. This work was supported by NASA grant NNX08AC146 to the University of Colorado at Boulder. Some of the data presented in this paper were obtained from the Multimission Archive at the Space Telescope Science Institute (MAST). Support for MAST for non-HST data is provided by the NASA Office of Space Science via grant NNX09AF08G and by other grants and contracts. 

{\it Facility:} \facility{HST (COS),Swift,ARC }

\clearpage
\begin{deluxetable}{lccccc}
\tablecaption{Observation Summary \label{tab_obs}}
\tablewidth{0pt}
\tablecolumns{6}
\tablehead{
\colhead{Telescope} & \colhead{Instrument} & \colhead{Grating/Filter} & \colhead{Date (UT)} & \colhead{Time (UT)} & \colhead{T$_{exp}$ (sec)}  }
\startdata
APO & SPIcam & BVRI & 2012 Oct 02 & 01:40 & 360 -- 700 \\
APO & NIC-FPS & JHK & 2012 Oct 02 & 02:59  & 108 -- 180 \\ 
HST & COS & G130M &  2012 Oct 02 & 08:14 & 5126  \\
HST & COS & G160M &  2012 Oct 02 & 11:24 & 5683  \\
HST & COS & G230L &  2012 Oct 02 & 14:34 & 2839  \\
SWIFT & XRT & \nodata & 2012 Oct 02 & 08:35 & 2120 \\
SWIFT & UVOT & U & 2012 Oct 02 & 08:39 &  2112 \\
\enddata
\end{deluxetable}

\clearpage
\begin{deluxetable}{lc}
\tablecaption{Optical and NIR Photometry \label{tab_oir}}
\tablewidth{0pt}
\tablecolumns{2}
\tablehead{\colhead{Filter} & \colhead{Magnitude}} 
\startdata
U & 16.19$\pm$0.02\tablenotemark{a} \\
B & 17.143$\pm$0.007 \\
V & 16.886$\pm$0.004 \\
R & 16.581$\pm$0.003 \\
I & 16.166$\pm$0.003 \\
J & 15.55$\pm$0.02 \\
H & 15.18$\pm$0.03 \\
K & 14.89$\pm$0.05 \\
\enddata
\tablenotetext{a}{Swift uvotsource magnitude calibrated in the Vega system.}
\end{deluxetable}

\clearpage
\begin{deluxetable}{ccc}
\tablecaption{Black Hole Mass and Distance Limits from UV Continuum Fits\label{tab_disk}}
\tablewidth{0pt}
\tablecolumns{3}
\tablehead{\colhead{$M_{BH}$} & \colhead{$d (i=55^{\circ}$)} & \colhead{$d (i=0^{\circ}$)} \\
\colhead{(\msun)} & \colhead{(kpc)} & \colhead{(kpc)} }
\startdata
3.0 & $\leq2.0$ & $\leq2.6$ \\
5.0 & $\leq2.8$ & $\leq3.7$ \\
7.0 & $\leq3.5$ & $\leq4.6$ \\
9.0 &$\leq4.1$ & $\leq5.5$ \\
12.0 &$\leq5.0$ & $\leq6.6$ \\
15.0 & $\leq5.8$ & $\leq7.7$ \\
\enddata
\tablecomments{For $i = 55^{\circ}$ and $\dot{m} \leq 0.05 \dot{m}_{Edd}$, equation (2) reduces to $d \leq 0.955 m^{2/3}$. For $i = 0^{\circ}$, $d \leq 1.262 m^{2/3}$.}
%\tablecomments{Model fits assume $\dot{m} = 0.05 \dot{m}_{Edd}$, $M_{2} = 0.3$~\msun, $P_{orb} = 3.23$~hr, $i = 55^{\circ}$, $r_{in} = 1.23 r_{g}$, $r_{out} = 0.6 R_{L_{1}}$. The third column shows the effect of decreasing the binary inclination on the distance upper limits.  }
\end{deluxetable}

\clearpage
\begin{deluxetable}{cccc}
\tablecaption{Spectral Energy Distribution \label{tab_sed}}
\tablewidth{0pt}
\tablecolumns{4}
\tablehead{
\colhead{Band} & \colhead{Instrument} & \colhead{$\log(\nu)$} & \colhead{$\log(\nu F_{\nu}$)}  \\
& & (Hz) & (ergs~cm$^{-2}$~s$^{-1}$)}
\startdata
X-ray & Swift & 16.87 -- 18.29 & -10.13 -- -9.86\tablenotemark{a} \\
FUV & COS & 15.41 & -9.83 \\
FUV & COS & 15.37 & -9.94 \\
FUV & COS & 15.35 & -9.96 \\
FUV & COS & 15.31 & -10.02 \\
FUV & COS & 15.24 & -10.11 \\
FUV & COS & 15.20 & -10.16 \\
NUV & COS & 15.12 & -10.28 \\ 
NUV & COS & 15.01 & -10.42 \\
U & UVOT & 14.93 & -10.49  \\
B & Spicam & 14.83 & -10.66 \\
V & Spicam & 14.74 & -10.90 \\
R & Spicam & 14.66 & -11.01 \\
I & Spicam &  14.56 & -11.14 \\
J & NIC-FPS & 14.38 & -11.49 \\ 
H & NIC-FPS & 14.26 & -11.68 \\
K & NIC-FPS & 14.14 & -11.96 \\
\enddata
\tablenotetext{a}{The X-ray data plotted is the model power law component of a powerlaw * phabs fit in xspec.} 
\tablecomments{The optical-NIR data have been dereddened assuming E(B--V) = 0.45. The X-ray data have been corrected for interstellar absorption assuming $N_{H} = 2.1\times10^{21}$~cm$^{-2}$.}
\end{deluxetable}

\clearpage
\pagestyle{empty}

\clearpage
\pagestyle{empty}

\clearpage
\pagestyle{empty}

\clearpage
\pagestyle{empty}

\clearpage
\pagestyle{empty}

\clearpage
\pagestyle{empty}

\end{document}